\documentstyle[prl,aps,psfig]{revtex}
\newcommand{\be}{\begin{eqnarray}}
\newcommand{\ee}{\end{eqnarray}}

\begin{document}
\twocolumn[\hsize\textwidth\columnwidth\hsize
           \csname @twocolumnfalse\endcsname
\title{Short time dynamics with initial correlations}
\author{K. Morawetz$^1$, M. Bonitz$^1$, V. G. Morozov$^2$, G. R{\"o}pke$^1$, D. Kremp$^1$}
\address{$^1$Fachbereich Physik, University Rostock, D-18055 Rostock,
Germany\\
$^2$Department of Physics, Moscow Institute
RAE, Vernadsky Prospect 78, 117454, Moscow, Russia}
\maketitle
\begin{abstract}
The short-time dynamics of correlated systems is strongly influenced by
initial correlations giving rise to an additional collision integral in the
non-Markovian kinetic equation. Exact cancellation of the two integrals is
found if the initial state is thermal equilibrium which is an important
consistency criterion. Analytical results are given for the time evolution
of the correlation energy which are confirmed by comparisons with
molecular dynamics simulations (MD).
\end{abstract}
\vskip2pc]

Although the Boltzmann kinetic equation is successfully applied to
many problems in transport theory, it has some serious shortcomings
\cite{SLM96}.
Among these, the Boltzmann equation cannot be used
on short time scales, where
memory effects are important~\cite{KBKS97,B99}.
In such situations, a frequently used non-Markovian kinetic equation
is the so-called Levinson equation~\cite{L65,L69}. One remarkable feature
of this equation is that it describes the formation of correlations in good
agreement with molecular dynamics simulations~\cite{MSL97a}.
Nevertheless, the Levinson equation is incomplete by two reasons:
(i) It does not include
correlated initial states. (ii) When the evolution of the system
starts from the equilibrium state,
the collision integral does not vanish, but gives rise to
spurious time evolution. The latter point has been addressed by
Lee et al.~\cite{Fu70} who clearly show that from initial
correlations there must appear terms in the kinetic
equation which ensure that the collision integral vanishes
in thermal equilibrium.

The aim of this letter is to derive the contributions from initial
correlations to the non-Markovian Levinson equation within perturbation
theory. We will restrict ourselves to the Born approximation
which allows us to present
the most straightforward derivation. The inclusion of higher order
correlations can be found in \cite{B99,D84,SKB99,MR99}.
The effect of initial correlations becomes particularly transparent from
our analytical results which may also serve as
a bench-mark for numerical simulations.

The outline of this Letter is as follows. First we give
the general scheme of inclusion of initial correlations into
the Kadanoff and Baym equations
in terms of the density fluctuation function.
We show that initial correlations enter the kinetic equation as
self-energy corrections and meanfield-like
contributions in terms of the initial two-particle
correlation function.
An analytical expression for the time dependent correlation
energy of a high temperature plasma is presented and then
compared with molecular dynamics simulations.

To describe density fluctuations we start with the causal density--density
correlation function \cite{KB62}
\begin{equation}\label{defL}
L(121'2')=G_2(121'2')-G(11')G(22'),
\end{equation}
where $1$ denotes cumulative variables $(x_1,t_1,..)$.
$G(1,2)={1\over i} \langle T \Psi(1)\Psi(2)^+\rangle$
 and $G_2(121'2')$ are the one- and two-particle
causal Green's functions. Their dynamics follows
the Martin- Schwinger hierarchy
\be
\lefteqn{
\left [i\hbar \frac{\partial}{\partial
t_1}+\frac{(\frac{\hbar}{i}\nabla_1)^2}
{2m} -\Sigma_H(1) \right ]G(1,1')=\delta(1-1')
}
\nonumber\\
&&\qquad\qquad\qquad\qquad 
+\int d3 V(1,3) L(1,3,1',3^+),
\nonumber\\
\lefteqn{
\left [i \hbar\frac{\partial}{\partial
t_1}+\frac{(\frac{\hbar}{i}\nabla_1)^2}
{2m} \right ]G_2(121'2')=
}
\nonumber\\
&&\qquad \qquad 
\delta(1-1') G(2,2')-\delta(1-2')G(2,1')\nonumber\\
&&\qquad\qquad\qquad 
+\int d3 V(1,3)
G_3(1,2,3,1',2',3^+),
\label{hier}
\ee
where $V(1,2)$ is the interaction amplitude and
 $\Sigma_H(1)=\int d2 V(1,2)G(2,2^+)$
is the Hartree self-energy.

Using for $G_3$ the polarization approximation
\begin{eqnarray}\label{g3}
G_3(1231'2'3')=G(11')G(22')G(33') +\nonumber\\
G(11')L(232'3')+G(22')L(131'3')+G(33')L(121'2'),
\end{eqnarray}
leads to a closed equation for $L$
which is conveniently rewritten as integral equation
\be
&&L(1,2,1',2')=L_0(1,2,1',2')-G_H(1,2')G(2,1')\nonumber\\
&&+\int d4 G_H(1,4)G(4,1')\int d3
V(4,3) L(2,3,2',3^+)
\label{L}
\ee
where $(G^R_{H})^{-1}$ denotes the l.h.s. of the first equation (\ref{hier})
and we have taken into account the boundary condition
\be
(G^R_{H})^{-1} L_0=0.
\label{grm1}
\ee
In the case that all times in (\ref{L}) approach $t_0$, the
right-hand side vanishes except
$L_0$ which represents, therefore, the contribution from initial correlations.
They propagate in time according to the solution of (\ref{grm1}) \cite{SKB99}
\be
&&L_0(121'2')=\int dx_1 dx_2 dx_1' dx_2'
G^R_{H}(1,x_1t_0)G^R_{H}(2,x_2t_0)
\nonumber\\&&\times L_{00}(x_1,x_2,x_1',x_2',t_0)
G^A_{H}(1',x_1't_0)G^A_{H}(2',x_2't_0).
\label{lrr}
\ee
Here $L_{00}$ is the initial two-particle correlation
function.

\begin{figure}
\centerline{\psfig{file=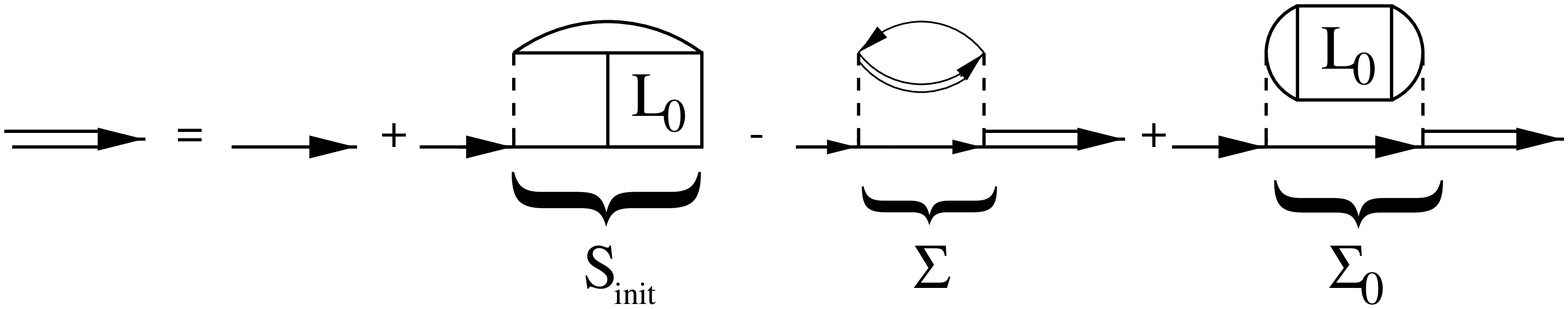,height=2cm,width=8cm,angle=0}}
\caption{The Dyson equation including density fluctuation up to 
second Born approximation.
Besides the initial correlation term ${\cal S}_{\rm init}$ discussed
in \protect\cite{D84,SKB99}, a new type of self energy $\Sigma_0$
appears which is induced by
initial correlations.
Since the latter one contains interaction by itself,
this term is of next order Born approximation.}
                      \label{2}
\end{figure}

Inserting (\ref{L}) into the first equation of (\ref{hier}) and
restricting to the Born approximation
we obtain for the causal function [$G_{HF}^{-1}(1,2)=G_H^{-1}(1,2)+V(1,2) G^<(1,2)$]
\be
&&G_{HF}^{-1}(1,3)G(3,2)=\delta(1-2)+{\cal S}_{\rm
init}(1,2)\nonumber\\
&&\qquad +\int\limits_{\cal C} d4 \left\{\Sigma_0(1,4)+\Sigma(1,4)\right\}G(4,2),
\label{dyson}
\ee
where the integration is performed along the Keldysh contour ${\cal C}$
with the
self-energy in Born approximation
\be
&&\Sigma(1,2)=\nonumber\\
&&\int d3 d5 V(1,3)
G_H(1,2) V(2,5)G_H(3,5^{++}) G(5,3^+).
\label{Born}
\ee
Two new terms appear due to initial correlations
\be
&&\Sigma_0(1,2)=\int d3 d5 V(1,3) G_H(1,2) V(2,5)
L_0(3,5,3^+,5^{++}),\nonumber\\
&&{\cal S}_{\rm init}(1,2)=\int d3 V(1,3)
L_0(1,3,2,3^+).
\label{dyson1}
\ee
The integral form of (\ref{dyson}) is given in figure \ref{2} from
which the
definitions (\ref{dyson1}) are obvious.

The equation for the retarded Green's function
$G^R(1,2)=-i \Theta(t_1-t_2) (G^>(1,2)+G^<(1,2))$, where
$G^<(1,2)=\langle\Psi^+(2)\Psi(1)\rangle$ and
$G^>(1,2)=\langle\Psi(1)\Psi^+(2)\rangle$, is
derived from (\ref{dyson}) as
\be
(G_{HF}^{-1}-\Sigma_0^R-\Sigma^R) G^R=\delta(1-2)+{\cal S}_{\rm
init}^R(1,2)
\label{gr}
\ee
and leads to the Kadanoff-Baym equation
\be
&&(G_{HF}^{-1}-\Sigma^R)G^<-G^<(G_{HF}^{-1}-\Sigma^A)\nonumber\\
&&=(\Sigma+\Sigma_0)^<G^A-G^R
(\Sigma+\Sigma_0)^<+{\cal S}_{\rm init}-{\cal S}_{\rm init}^*.
\label{ret}
\ee
Using the generalized Kadanoff-Baym ansatz
\cite{LSV86}
\be
G^<(t_1,t_2)=iG^R(t_1,t_2)\rho(t_2)-i\rho(t_1)G^A(t_1,t_2)
\label{GKB}
\ee
we obtain the kinetic equation for the reduced
density matrix $\rho(t)=G^<(t,t)$ 
\be
&&{\partial \over \partial t}{\rho}(k,t)={\cal I}(k,t)+{\cal I}_0(k,t)+{\cal I}_1(k,t)
\label{ki}
\ee
with
\be
&&{\cal I}(k,t)={2 \over \hbar^2} {\rm Re}\int\limits_{t_0}^t dt_1\int {d q
dp\over (2 \pi
\hbar)^6} V^2(q) \nonumber\\
&&\times
G^R(t,t_1,k-q)G^A(t_1,t,k) G^R(t,t_1,p+q)G^A(t_1,t,p)
\nonumber\\
&&\times
\left [\rho(t_1,k-q)\rho(t_1,p+q)(1-\rho(t_1,p))(1-\rho(t_1,k))\right .
\nonumber\\
&&\left . -\rho(t_1,k)\rho(t_1,p)(1-\rho(t_1,p+q))(1-\rho(t_1,k-q))\right ],
\label{kin1}\\
&&{\cal I}_0(k,t)={2\over \hbar} {\rm Im}\int {d q dp\over (2 \pi \hbar)^6} V(q)
\nonumber\\
&&\times
G^R(t,t_0,k-q)G^A(t,t_0,k) G^R(t,t_0,p+q)G^A(t,t_0,p) \nonumber\\
&&\times
\langle{p-k\over 2}+q|L_{00}(p+k,t_0)|{p-k\over 2}\rangle,
\label{kin3}\\
&&{\cal I}_1(k,t)={2 \over \hbar^2} {\rm Re} \int\limits_{t_0}^t dt_1\int {d q \over
(2 \pi
\hbar)^3} {\cal L}_0(q,t,t_1) V^2(q)
\nonumber\\
&&\times
G^R(t,t_1,k-q)G^A(t_1,t,k) 
\left [\rho(t_1,k-q) -\rho(t_1,k) \right ]
\label{kin2}
\ee
where $L_{00}(x_1,x_2,x_3,x_4)=<x_1-x_2|L_{00}({x_1+x_2\over 2}-{x_3+x_4\over 2})|x_3-x_4>$ and 
${\cal L}_0(q,t,t')=\int dx {\rm e}^{-iq x}<{x\over 2}|L_0(0)|{x\over 2}>$.
We like to note that the equation (\ref{ki}) is valid up to second order gradient expansion in the spatial coordinate. This variable has to be added simply in all functions and on the left side of (\ref{ki}) the standard meanfield drift appears.

The first part (\ref{kin1}) is just the precursor of the Levinson
equation in
second Born approximation $\sim V^2$.
The term (\ref{kin2})
coming from $\Sigma_0$ leads to
corrections to the third Born approximation since it is $\sim V^2{\cal L}_0$.
A more general discussion of higher-order
correlation contribution
within the T-matrix
approximation can be found in \cite{KBKS97,MR99} and of general initial conditions in \cite{ZMR96a}.
The second part (\ref{kin3}) following from ${\cal S}$
gives just the correction to the Levinson equation,
which will guarantee the cancellation of the
collision integral for an equilibrium
initial state. Recently the analogous term in the collision integral has been
derived by other means \cite{SKB99}.

Multiplying the kinetic equation equation (\ref{ki}-\ref{kin3})
with a momentum function
$\phi(k)$ and integrating over k, one derives the balance equations
\be
\langle\dot \phi(k)\rangle=\int {d k \over (2 \pi \hbar)^3} \phi(k) {\cal I}+\int
{d k \over
(2 \pi \hbar)^3} \phi(k) {\cal I}_0.
\label{e}
\ee
For the standard collision integral follows
\be
&&\langle\phi(k) {\cal I}\rangle={1\over \hbar^2}{\rm Re}\int {d k d q d p\over (2 \pi
\hbar)^9}
\int_{t_0}^t dt_1 V^2(q)\nonumber\\&&
\times  G^R(t,t_1,k-q)G^A(t_1,t,k)
 G^R(t,t_1,p+q)G^A(t_1,t,p)
\nonumber\\&&\times
 \rho(t_1,k-q)\rho(t_1,p+q)
(1-\rho(t_1,p))(1-\rho(t_1,k))
\nonumber\\&&\times
\biggl\{\phi(k)+\phi(p)-\phi(k-q)-\phi(p+q)\biggr\},
\label{con}
\ee
from which it is obvious that density and momentum
($\phi=1,k$) are conserved, while a change of kinetic energy $\phi=k^2/2m$
is induced
which exactly compensates the two-particle correlation energy and, therefore,
assures total energy conservation of a correlated plasma \cite{M94}.
Initial correlations, Eq. (\ref{kin3}), give rise to additional contributions to the balance
equations \cite{B99,SKB99}. We get  
\be
&&\langle\phi(k) {\cal I}_0\rangle
={1\over 4 \hbar}\int {d k d q d p\over (2 \pi
\hbar)^9}
V(q)
\nonumber\\&&
\times\left (\langle {p-k\over 2}+q|L_0(p+k)|{p-k\over 2}\rangle-c.c. \right )
\nonumber\\&&
\times \left \{\phi(k)+\phi(p)-\phi(k-q)-\phi(p+q)\right \}
\label{con0}
\ee
which keeps the density and momentum also unchanged and only a correlated energy is induced.
 The self-energy
 corrections from initial correlations which correct
 the next Born approximation, (\ref{kin2}), would lead to
 \be
 &&\langle\phi(k) {\cal I}_1\rangle={2 \over \hbar^2} {\rm Re}\int {d k d q 
\over (2 \pi
 \hbar)^6}
 \int\limits_{t_0}^t dt_1 V^2(q) {\cal L}_0(q,t,t_1)  \rho(t_1,k)
 \nonumber\\&&\times
 G^R(t,t_1,k-q)G^A(t_1,t,k)
 \biggl\{\phi(k-q)-\phi(k)\biggr\}
 \label{conc}
 \ee
 which shows that the initial correlations induce a flux
 besides an energy in order to equilibrate the correlations imposed initially
 towards the
 correlations developed during dynamical evolution if higher than $\sim V^2$ correlations are considered.

 We will consider in the following only second Born approximation
 $\sim V^2$ and have therefore to use from (\ref{gr})
\be
G^R(t_1,t_2,k)&\approx&-i\Theta(t_1-t_2) {\rm e}^{i {k^2\over 2
m \hbar}(t_2-t_1)},
\label{gr1}
\ee
and for $L_{00}$ the first Born approximation
\be
&&\left\langle{k-p\over 2}\left|L_{00}(k+p)\right|{k-p\over 2}-q\right\rangle
\nonumber\\
&&=-{{\cal P}\over \Delta \epsilon} V_0(q)\left\{\rho_0(k)\rho_0(p)
(1-\rho_0({k-q}))(1-\rho_0({p+q}))\right .\nonumber\\
&&\left . \,\,\,-(1-\rho_0(k))(1-\rho_0(p))
(\rho_0({k-q}))(\rho_0({p+q}))
\right\}.
\label{equi1}
\ee
where ${\cal P}$ denotes the principal value, $\Delta \epsilon={k^2 \over 2 m}+{p^2 \over 2 m}-{(k-q)^2 \over 2 m}-{(p+q)^2\over 2 m}$ and $\rho_0$ the initial Wigner distribution.
Then the explicit collision integral 
(\ref{kin1}) reads
\be
&&{\cal I}(k,t)=
{2 \over \hbar^2} \int_{t_0}^t dt_1\int {d q dp\over (2 \pi
\hbar)^6}
V^2(q)
\nonumber\\
&&\times
\cos{\biggl[\left ({k^2\over 2 m}+{p^2\over 2 m}-{(k-q)^2\over 2
m}-{(p+q)^2\over 2
m}\right ){(t-t_1)\over \hbar}\biggr]} \nonumber\\
&&\times
\left \{\rho(t_1,k-q)\rho(t_1,p+q)(1-\rho(t_1,p)-\rho(t_1,k))
\right .
\nonumber\\
&&\times
\left .-\rho(t_1,k)\rho(t_1,p)(1-\rho(t_1,p+q)-\rho(t_1,k-q))\right
\}
\label{stat}
\ee
and the new term due to initial correlations (\ref{kin3}) is
\be
&&{\cal I}_0(k,t)=
{2 \over \hbar^2} \int_{t_0}^t dt_1\int {d q dp\over (2 \pi
\hbar)^6}
V(q)V_0(q)
\nonumber\\
&&\times
\cos{\biggl[\left ({k^2\over 2 m}+{p^2\over 2 m}-{(k-q)^2\over 2
m}-{(p+q)^2\over 2
m}\right ){(t-t_1)\over \hbar}\biggr]} \nonumber\\
&&\times
\left \{\rho_0(k-q)\rho_0(p+q)(1-\rho_0(p)-\rho_0(k))
\right .
\nonumber\\
&&\times
\left .-\rho_0(k)\rho_0(p)(1-\rho_0(p+q)-\rho_0(k-q))\right
\}. 
\label{stat0}
\ee

To show the interplay between collisions and
correlations, we have calculated the initial two-particle
correlation function in the ensemble, where
the dynamical interaction $V(q)$ is replaced by some
arbitrary function $V_0(q)$. Therefore the initial state 
deviates from thermal equilibrium except
when $V(q)=V_0(q)$ and $\varrho(t_0)=\varrho_0$.

The additional collision term, ${\cal I}_0$, cancels
exactly
the Levinson collision term in the case that we have initially
the same interaction
as during the dynamical evolution ($V_0=V$)
and if the system starts from the
equilibrium $\rho(t)\equiv \rho_0$.
Therefore we have completed our task and derived a
correction of the Levinson
equation which ensures the cancellation of the
collision integral
in thermal equilibrium \cite{note}.

\begin{figure}
\centerline{\psfig{file=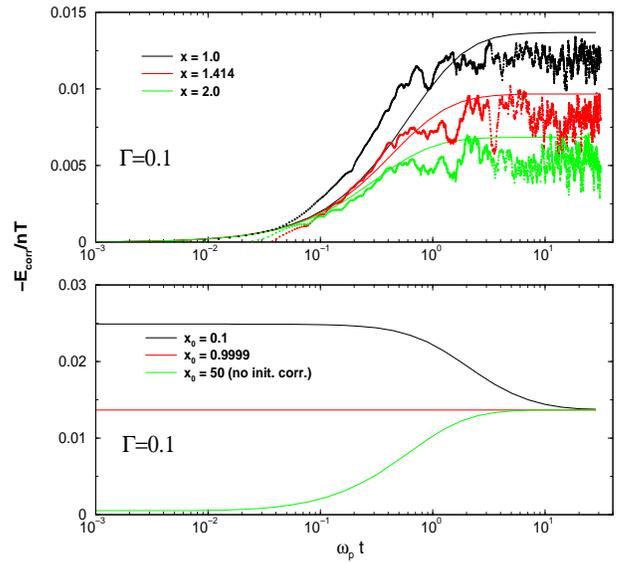,height=7.5cm,width=8cm,angle=-90}}
\caption{The formation of correlation energy
$-{\cal E}_{\rm corr}={\rm E}_{\rm total}-{\cal E}_{\rm init}-
{\cal E}_{\rm coll}={\cal E}_{\rm kin}$
in a plasma with Debye interaction $V_i$. The upper panel compares the
analytical results
(\protect\ref{state}) with MD simulations from
\protect\cite{GZ99} for
three different ratios of $\kappa_D$ to the inverse Debye length
$x=\kappa_D/\kappa$.
In the lower panel we compare theoretical predictions for the
inclusion of Debye initial correlations characterized by
$x_0=\kappa_0/\kappa$ where
$x=\kappa_D/\kappa=1$.}
                      \label{zwick3}
\end{figure}

On very short time scales we can neglect the change in the
distribution function. Assuming a Maxwellian initial  distribution with
temperature $T$ and neglecting degeneracy, we can calculate explicitly
the collision integrals and obtain analytical results.
We choose as a model interaction a Debye potential
$V_i(q)=4 \pi e^2\hbar^2 / [q^2+\hbar^2\kappa_i^2]$
with fixed
parameter $\kappa_i=\kappa_D$
and for the initial correlations $\kappa_i=\kappa_0$.
We obtain for the change of kinetic energy on short times from
(\ref{con}) and
(\ref{con0})
\be
{\partial \over \partial t}E_{\rm kin}(t)={\cal E}[V(q)^2](t)-{\cal
E}[V_0(q)V(q)](t),
\label{26}
\ee
which can be integrated \cite{MSL97a} to yield
\be
E_{\rm kin}(t)=E_{\rm total}-E_{\rm init}(t)-E_{\rm coll}(t).
\label{et}
\ee
For the classical limit
we obtain explicitly the time dependent kinetic energy
\be
&&{E_{\rm coll}(t)\over nT}
=-{\sqrt{3} \Gamma^{3/2}\over 4 x}
\partial_y (y {\cal F}(y))_{y=x \tau},
\label{state}
\ee
where ${\cal F}(y)=1-{\rm e}^{y^2}{\rm erfc}(y),$
$\tau= t \omega_p/\sqrt{2}$, $x=\kappa_D/\kappa$ and
$\kappa^2=4\pi e^2 n/T=\omega_p^2 T/m$.
The plasma parameter is given as usually by
$\Gamma={e^2\over a_eT}$, where
$a_e=({3\over 4\pi n})^{1/3}$ is the Wigner-Seitz radius.

In Fig. \ref{zwick3}, upper panel, we compare the analytical results of
(\ref{state})
with MD simulations \cite{GZ99} using the Debye
potential $V_i$ as bare interaction.
The evolution of kinetic energy is
shown for three different ratios $x$. The agreement between theory and
simulations is quite satisfactory, in particular, the short time
behavior
for $x = 2$. The stronger initial increase of kinetic energy
observed in the simulations at $x=1$ may be due to the
finite size of the
simulation box which could more and more affect the results for
increasing range of the interaction. 

Now we include the initial correlations choosing the
equilibrium expression (\ref{equi1}) which leads to
\be
{E_{\rm init}(t)\over nT}
&=&-{\sqrt{3} \Gamma^{3/2}\over {2 (x_0^2-x^2)}}
\left[ x {\cal F}(x\tau) - x_0 {\cal F}(x_0 \tau)\right],
\label{dyne0}
\ee
where $x_0=\kappa_0/\kappa$ characterizing the strength of the initial
Debye correlations (\ref{equi1}) with the Debye potential $V_0$ which
containes $\kappa_0$ instead of $\kappa_D$. Besides the
kinetic energy (\ref{dyne0}) from initial correlations, the total energy
$E_{\rm total}$ (\ref{et}) now includes the initial correlation energy
which can be calculated
from the long time limit of (\ref{state}) leading to
\be
{E_{\rm total}\over nT}={\sqrt{3} \Gamma^{3/2}\over 2 (x+x_0)}.
\ee
The result (\ref{et}) is seen in Fig. \ref{zwick3}, lower panel. We
observe that if the initial correlation is characterized by a potential
range
larger than the Debye screening length, $x_0 < 1$, the initial state is
over--correlated, and the correlation energy starts at a higher absolute
value than without initial correlations relaxing towards the correct
equilibrium value.
If, instead, $x_0 = 1$ no change of correlation energy is observed, as expected.
Similar trends have been observed in numerical solutions
\cite{SKB99}.

In summary, in this Letter initial correlations are investigated within
kinetic theory. Explicit correction terms appear on every level of
perturbation theory correcting the non-Markovian kinetic equation
properly in a way that the collision integral
vanishes if the evolution starts from a correlated equilibrium state.
Furthermore, the conservation laws of a correlated plasma are proven
including the contributions from initial correlations.
It is shown that besides the appearance of
correlation energy a correlated flux appears
if higher than Born correlations are considered.

Deriving analytical formulas for high temperature plasmas
allowed us to investigate the time dependent formation of the correlation
energy and the decay of initial correlations.
 The comparison with
 molecular dynamics simulations is found to be satisfactorily.
 Including initial correlations the cases of over- and
 under-correlated initial
 states are discussed. While starting from equilibrium the
 correlation
 energy does not change, for over- and under-correlated states the equilibrium
 value is approached after a time of the order of the inverse plasma frequency.

The many interesting discussions with Pavel Lipavsk\'y, V\'aclav \v
Spi\v cka
and D. Semkat are gratefully acknowledged.
G. Zwicknagel is thanked for providing simulation data prior to publication.


\begin{thebibliography}{10}

\bibitem{SLM96}
V. {\v S}pi{\v c}ka, P. Lipavsk{\'y}, and K. Morawetz, Phys. Lett. A {\bf 240},
   160  (1998).

\bibitem{KBKS97}
D. Kremp, M. Bonitz, W. Kraeft, and M. Schlanges, Ann. of Phys. {\bf 258},  320
   (1997).

\bibitem{B99}
M.~Bonitz, {\em Quantum Kinetic Theory} (Teubner, Stuttgart, 1998).

\bibitem{L65}
I.~B. Levinson, Fiz. Tverd. Tela Leningrad {\bf 6},  2113  (1965).

\bibitem{L69}
I.~B. Levinson, Zh. Eksp. Teor. Fiz. {\bf 57},  660  (1969), [Sov. Phys.--JETP
  {\bf 30}, 362 (1970)].

\bibitem{MSL97a}
K. Morawetz, V. {\v S}pi{\v c}ka, and P. Lipavsk{\'y}, Phys. Lett. A {\bf 246},
   311  (1998).

\bibitem{Fu70}
D. Lee, S. Fujita, and F. Wu, Phys. Rev. A {\bf 2},  854  (1970).

\bibitem{D84}
P. Danielewicz, Ann. Phys. (NY) {\bf 152},  239  (1984).

\bibitem{SKB99}
D. Semkat, D. Kremp, and M. Bonitz, Phys. Rev. E {\bf 59},  1557  (1999).

\bibitem{MR99}
V.G. Morozov, G. R\"opke, Ann. Phys. (NY) submitted.

\bibitem{KB62}
L.~P. Kadanoff and G. Baym, {\em Quantum Statistical Mechanics} (Benjamin, New
  York, 1962).

\bibitem{LSV86}
P. Lipavsk{\'y}, V. {\v S}pi{\v c}ka, and B. Velick{\'y}, Phys. Rev. B {\bf
  34},  6933  (1986).

\bibitem{ZMR96a}
D.~N. Zubarev, V. Morozov, and G. R{\"o}pke, {\em Statistical Mechanics of
  Nonequilibrium Processes} (Akademie Verlag, Berlin, 1997), Vol.~2.

\bibitem{M94}
K. Morawetz, Phys. Lett. A {\bf 199},  241  (1995).

\bibitem{note}
It is interesting to note
 that the corrections to the next Born
 approximation (\ref{kin2}) due to initial correlations is of the type
 found in
 impurity scattering. Therefore the initial correlations higher than
 $\sim V^2$
 are governed by another type of dynamics
 than the build up of correlations
 involved
 in ${\cal I}$ and ${\cal I}_0$.


\bibitem{GZ99}
G. Zwicknagel, Contrib. Plasma Phys. {\bf 39} (1999) 1-2,155,
and private communications

\end{thebibliography}

\end{document}